\documentclass[aps,prl,twocolumn]{revtex4-1}
\usepackage{graphicx}
\usepackage{epsfig}
\usepackage{bm}
\usepackage{amsmath}
\usepackage{amssymb}
\usepackage{wasysym}
\usepackage{epstopdf}

\begin{document}

%=============== TITLE ================

\title{Flow reversals in thermally driven turbulence}

\author{Kazuyasu Sugiyama$^{1,3}$, Rui Ni$^{2}$,  Richard J. A. M. Stevens$^1$, Tak Shing Chan$^{1,2}$, Sheng-Qi Zhou$^2$,
Heng-Dong Xi$^2$,
  Chao Sun$^{1,2}$, Siegfried Grossmann$^4$, Ke-Qing Xia$^2$, and Detlef Lohse$^1$}

\affiliation{$^{1}$Physics of Fluids Group,
Faculty of Science and Technology, Impact and Mesa$^+$ Institutes \& Burgers Center for Fluid Dynamics, University of Twente, 7500AE Enschede, The Netherlands\\
$^2$ Department of
Physics, The Chinese University of Hong Kong, Shatin, Hong Kong,
China\\
$^3$ Department of Mechanical Engineering, School of Engineering, The University of Tokyo, Tokyo, Japan\\
$^4$ Fachbereich Physik, Renthof 6, D-35032 Marburg, Germany
}
\date{\today}

\begin{abstract}
 
We analyze the reversals of the large scale flow in Rayleigh-B\'enard convection both through particle image velocimetry flow visualization and direct numerical simulations (DNS) of the underlying Boussinesq equations in a (quasi) two-dimensional, rectangular
 geometry of  aspect ratio 1. For medium Prandtl number there is a diagonal large scale convection roll and two smaller secondary rolls in the two remaining corners diagonally opposing each other. These corner flow rolls play a crucial role for the large scale wind reversal: They grow in kinetic energy and thus also in size thanks to plume detachments from the boundary layers up to the time that they take over the main, large scale diagonal flow, thus leading to reversal. Based on this mechanism we identify a typical time scale for the reversals. We  map out the Rayleigh number vs Prandtl number phase space and find that the occurrence of reversals very sensitively depends on these parameters.

\end{abstract}

\maketitle

Spontaneous flow reversals occur in various buoyancy driven fluid dynamical systems, including large scale flows in the ocean, the atmosphere, or the inner core of stars or the earth, where such reversals are associated with the reversal of the magnetic field. The paradigmatic example for buoyancy driven flow is the Rayleigh-B\'enard system, i.e., a fluid-filled cell heated from below and cooled from above, see e.g.\ the recent reviews \cite{ahl09,loh10}. In this system flow reversals have been detected and statistically analyzed, mainly through measurements of the temperature at one \cite{sre02} or several points \cite{bro06,bro07} in the flow or at the walls and more recently through flow visualization with particle image velocimetry (PIV) \cite{xi07,xi08c}. Various models have been developed to account for the reversals, either of stochastic nature \cite{ben05,BA08a} or based on simplifying (nonlinear) dynamical equations \cite{fon05,res06}, which show chaotic deterministic behavior. Most of the experimental studies have so far been done in a cylindrical cell, where the complicated three-dimensional dynamics of the convection role (see e.g. \cite{xi09,bro09} and section VIII of \cite{ahl09}) complicates the identification of the reversal process.

In the present paper, we restrict ourselves to the study of flow reversals in (quasi) two-dimensional (2D) rectangular geometry:
 experimentally to RB convection in a flat cell and numerically to DNS of the two-dimensional Oberbeck-Boussinesq equations,
for which reversals have been observed already in \cite{sch02}. This approach offers three advantages: (i) The flow reversal in quasi-2D is less complicated than in 3D (and therefore of course may be different); (ii) the visualization of the full flow is possible; and (iii) a study of a considerable fraction of the Rayleigh number 
$Ra$  %$ = \beta g H^3 \Delta / \nu \kappa$ 
vs Prandtl number $Pr$   %$ = \nu / \kappa$ 
phase space becomes numerically feasible.

The experiments were made in rectangular, quasi-2D cells \cite{xia03}. To extend the range of Ra, two cells of nearly identical geometry are used.  The larger (smaller) cell has a horizontal cross section of 24.8$\times$7.5 (12.6$\times$3.8) cm$^2$, and the heights of the larger ( smaller) cell is  $H=$ 25.4 cm ($H=$ 12.6 cm), 
giving an aspect ratio 
$\Gamma \approx 1$ in the plane of the main flow (and an aspect ratio of about 0.3 in the direction perpendicular to it). The fluid is water with a mean temperature of 28$^o$C, corresponding to Pr=5.7. For direct visualization of flow reversals,  PIV is used for a few selected Ra. The PIV measurement in this system has also been described previously \cite{xia03}. To study the statistical properties of the reversals over long time periods, we measure the temperature contrast $\delta T$ between two thermal probes embedded respectively in the left and right sides of either the top or the bottom plates. Reversals of the upward going hot plumes and downward going cold plumes correspond to the switching between the right and left sides of the system, $\delta T$ therefore is indicative of reversals. 

The numerical code is based on a fourth-order finite-difference discretization of the incompressible Oberbeck-Bousinesq equations and has been described in \cite{sug09}. The grid resolution has been chosen according to the strict requirements as formulated in \cite{ste10}. As in experiment also the numerical flow is wall-bounded, i.e., we use no-slip boundary conditions at all solid boundaries: $\bf{u}=0$ at the top ($z=H$) and bottom ($z=0$) plates as well as on the side walls $x=0$ and $x=H$. For the temperature at the side walls heat-insulating conditions are employed and $T_b-T_t = \Delta$ is the temperature drop across the whole cell. Times are given in multiples of the large eddy turnover time 
$t_E$, defined by 
$t_E := 4\pi / \left<  |\omega_c (t)| \right> $, where  $\omega_c$ denotes 
the center vorticity.

\begin{figure*}
\centering
\includegraphics[width=0.99\textwidth]{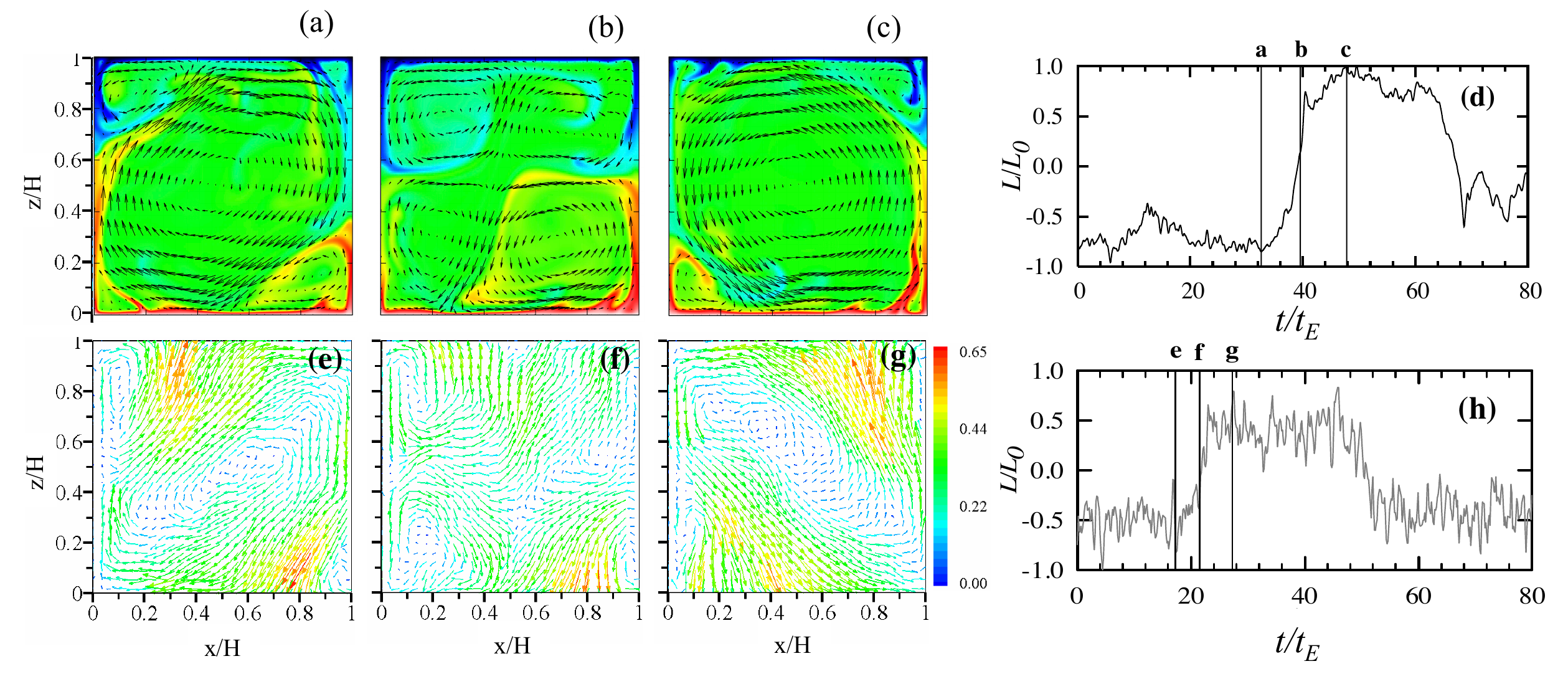}
\caption
{(Color online) Top panel: Snapshots of the temperature (color) and velocity (arrows) 
field and time trace of angular momentum from numerical simulations ($Ra=10^8$ and $Pr=4.3$). Bottom panel: Snapshots of the velocity field and time trace of angular momentum from experiment ($Ra=3.8\times 10^8$ and $Pr=5.7$). (a), (b), and (c) show the instantaneous dimensionless temperature $(T-T_t)/\Delta$ distribution. (d) shows the temporal change of the dimensionless angular momentum $L(t)/L_0$, where $L_0$ is the maximum of the absolute value of  $L$. The positive and negative signs indicate the anti-clockwise and clockwise circulations, respectively. (e), (f), and (g) show the PIV-measured instantaneous velocity field and (h) the normalized instantaneous angular momentum $L(t)/L_0$. The color bar indicates the %respective 
magnitude of the %temperature and 
velocity (in unit of cm/sec). 
The snapshots in (a), (b), and (c) give numerical, and those in (e), (f), and (g) give experimental examples of the large scale circulation  before, during and after a reversal process, as indicated in (d) and (h) respectively. Note that (b) and (f) show clearly the key role played by the growth of the corner rolls in the reversal process. Corresponding movies are offered in the supplementary material.
}
\label{fig:snap_num}
\end{figure*}

We start by showing qualitative features of the reversal process using examples from both numerical simulations and experiments. Fig.\ref{fig:snap_num} shows snapshots of the temperature and velocity fields from DNS and those of the velocity field from experiment just before, during, and after the large convection roll reversal. Corresponding videos can be viewed from the supplementary materials. Visually, the reversal process can be easily detected. To automatize this we measure the {\it local} angular velocity at the center of the cell; however, with this method some plumes passing through the center can lead to erroneous reversal counting. A better way is to rely on a {\it global} quantity, e. g. the global angular momentum (which has been successfully used for reversal characterization in 2D Navier-Stokes turbulence \cite{hei06,sug09}). This is defined as ${L}(t)=\langle-(z-H/2)u_x({\bm x},t)+(x-H/2)u_z({\bm x},t)\rangle_V$, where $\langle...\rangle_V$ represents averaging over the full volume. The time dependence of $L$ from simulation and experiment, as shown respectively in Figs.\ref{fig:snap_num}(d) and (h), indeed nicely reveals the reversal through a sign-change.

From the movies corresponding to Fig.\ \ref{fig:snap_num} the basic role of the corner flows in the reversal process can be observed: While the main roll is diagonally orientated in the cell, smaller counter-rotating rolls develop in diagonally opposing corners. They are energetically fed by detaching plumes from the boundary layers trapped in the corner flows, leading to their growth. Once the two corner flows have reached a linear extension of $\approx H/2$ [Fig.\ref{fig:snap_num}(b) and moment (b) in Fig.\ref{fig:snap_num}(d) and Fig.~\ref{fig:snap_num}(f) and moment (f) in Fig.\ref{fig:snap_num}(h)], they 
destroy the main convection roll and establish another one circulating in opposite direction. 

%%%%%%%%%%%%%%%%%%%%%%%%%

\begin{figure}[htbp]
\centering
\includegraphics[width=0.45\textwidth]{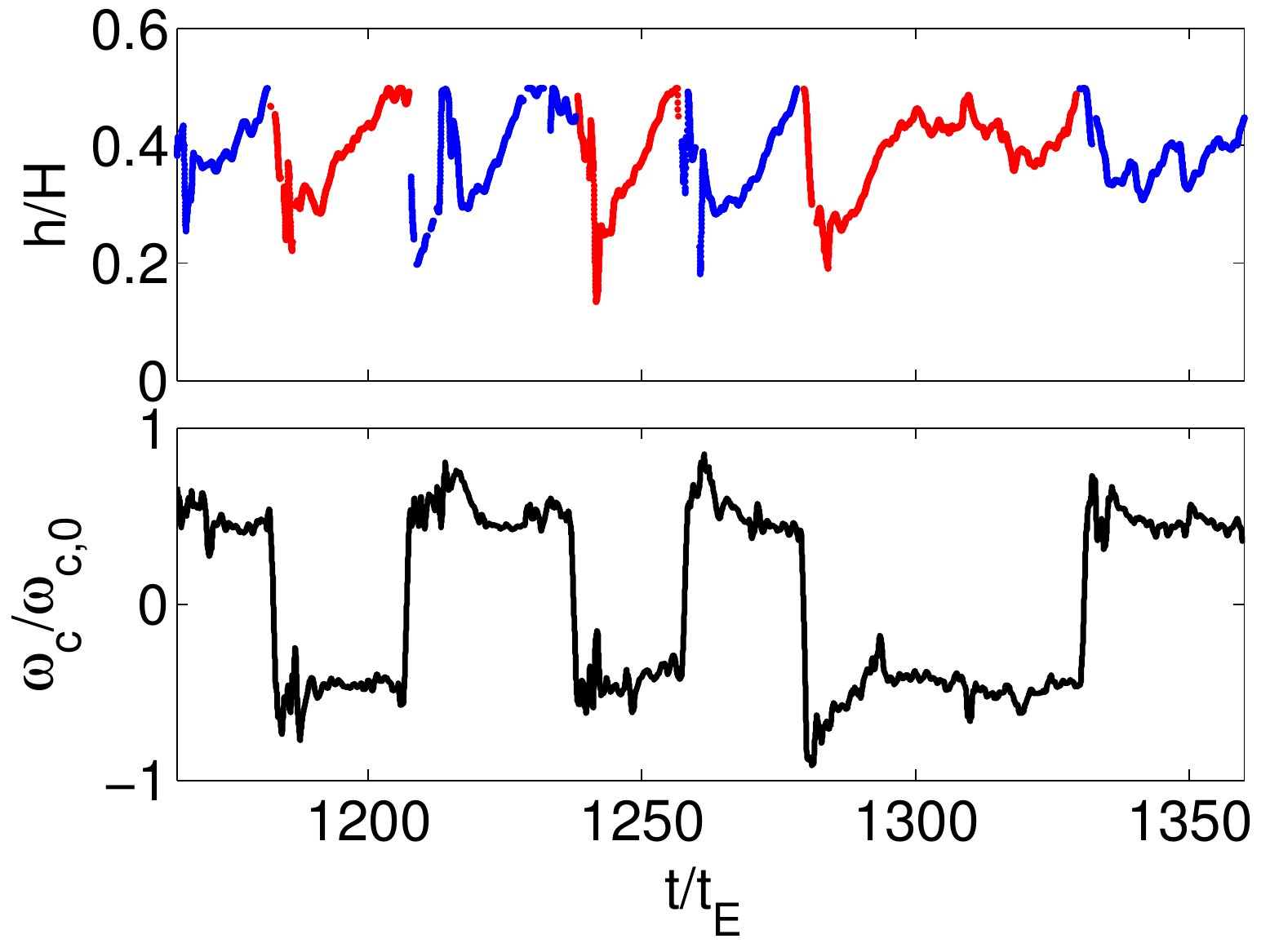}
\caption
{(Color online) 
Time series of the center vorticity $\omega_c(t)$ (rescaled by its maximum)  
(lower panel) and the heights h(t) of the lower
left (blue) and right (red) corner flows, revealing their approximate linear growth.
Not all growth processes need to lead to an immediate successful reversal, 
as seen for $t/t_E \approx 1300$.  
$Ra = 10^8$, $Pr= 4.3$.
}
\label{fig:fig2new}
\end{figure}

The heights $h(t)$ of the corner flows are
measured by
 first fitting the temperature profile at the respective sidewall with splines, and
then identifying the position of the steepest gradient of $T(z)$: From 
movies and snapshots we judge that this is an excellent measure for the height $h(t)$ of the
corner-flow. 
Time series of $h(t)$, together with the (rescaled) center vorticity $\omega_c(t)$
as quantitative measure of the strength of the large scale convection roll,  
 are shown in figure \ref{fig:fig2new}.
It is seen that after a reversal (as indicated by a sign change in $\omega_c(t)$) the 
respective corner-flow grows roughly linearly in time, before it reaches 
the half-height $h(t)/H\approx 1/2$
and breaks down, leading to flow reversal. 
However, 
the growth of the corner-flow need not always
lead to a reversal of the large scale convection roll: There are cases in which the corner-flow
looses energy due to some plume detachment from it, leading to full recovery 
of the large scale convection roll in its orginal direction
(e.g., at $t/t_E\approx 1300$ in figure \ref{fig:fig2new}). 
Also in experiment we have observed such unsuccessful build-ups of the corner-flow. 
Below we will try to quantify the energy gains and losses of the corner-flows.

\begin{figure}[htbp]
\centering
\includegraphics[width=0.45\textwidth]{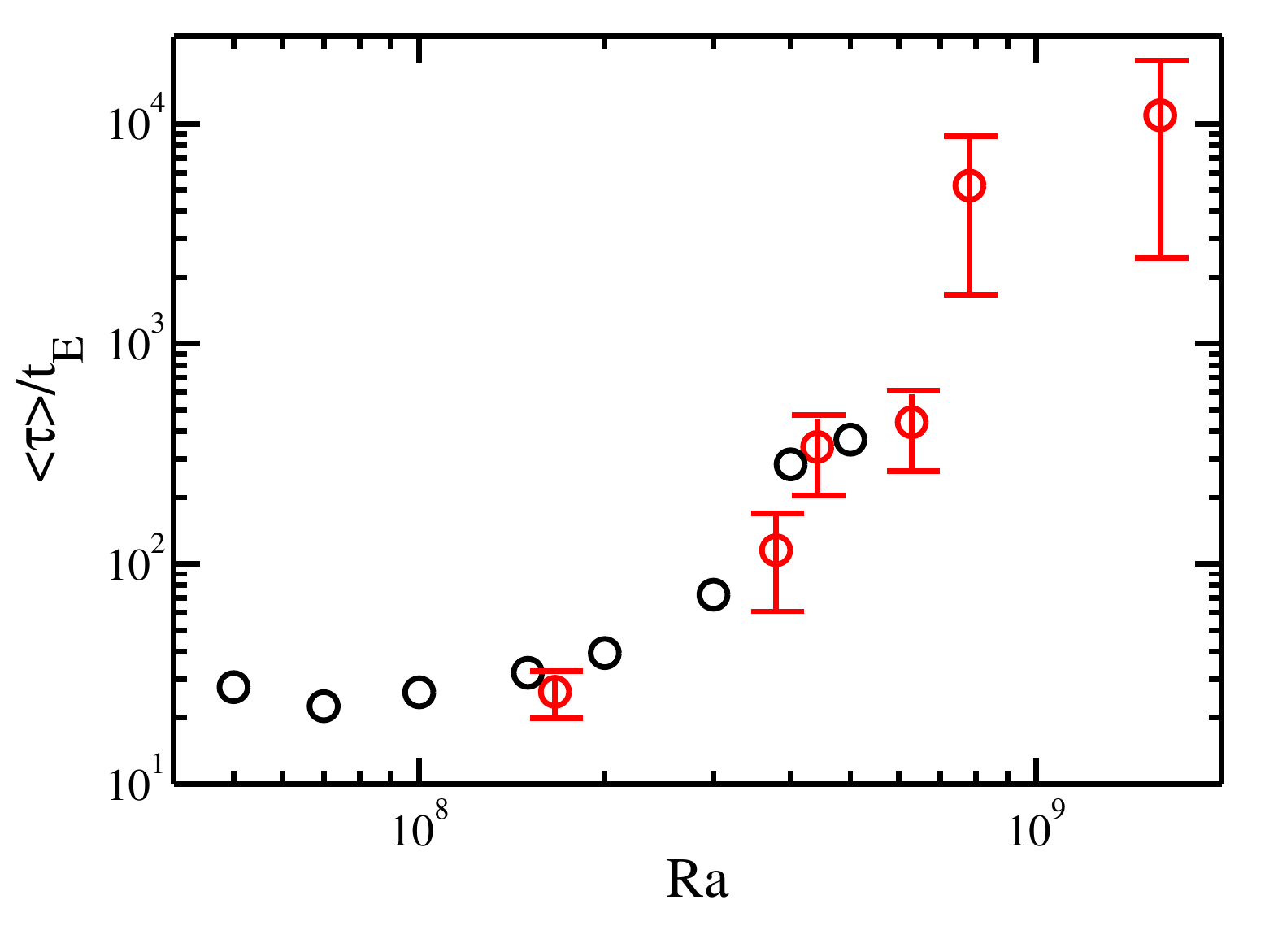} 
\caption
{(Color online) Log-log plot of the Ra-dependence of the mean time intervals $\left< \tau \right> $ between reversals, normalized in terms of
the large eddy turnover time $t_E$. 
Red symbols are from experiment and black ones from simulation. The error bar 
originates from the statistics of the reversals; 
for the numerical case it is smaller than the symbol size.
}
\label{fig:interval}
\end{figure}

The mean time interval $\left<\tau\right>$ 
%between two reversals is found to increase with increasing Ra, see
is shown in Fig.~\ref{fig:interval}. 
Firstly, we clearly see that experiment and simulation are in very good agreement.  
%for the experiment data point of  $Ra$ = 1.6 $\times$ 10$^9$, giving the extremely long measurement time (more than 10,000 turn over times). 
The  figure shows that 
 $\left< \tau \right>/ t_E $ at most weakly depends on Ra up to $Ra\approx 2 \times 10^8$, but
for larger Ra rapidly increases 
with increasing Ra, i.e., reversals occur less and less frequently.
 The numbers mean that 
there are only very few reversals: of order one per hour in the $Ra \sim 10^8$ range down to one 
within two days in the $Ra \sim 10^9$ range.
For 
Ra$\ge  5\times  10^8$ 
no reversals could be detected any more in our numerical simulations,
even for an averaging time of 2600 large eddy turnovers (see accompanying movie).
In experiment
two  reversals could still be observed at
Ra$= 1.6 \times  10^9$ (presumably due to the longer observation time in experiment
which goes beyond
10000 large eddy turnovers, corresponding to four days), 
but these also cease for larger Ra. 
%Note that the mean time between reversals expressed in the viscous time scale, $\left< \tau\right> / (H^2/\nu )$ (main figure), numerally increases less than the one expressed in large eddy turnover times $t_E$ (inset), as $t_E \sim (H^2/\nu) Re^{-1}$ and $Re\sim Ra^{0.62}$ in the 2D simulations \cite{sug09}. 

\begin{figure}[htbp]
\centering
\includegraphics[width=0.45\textwidth]{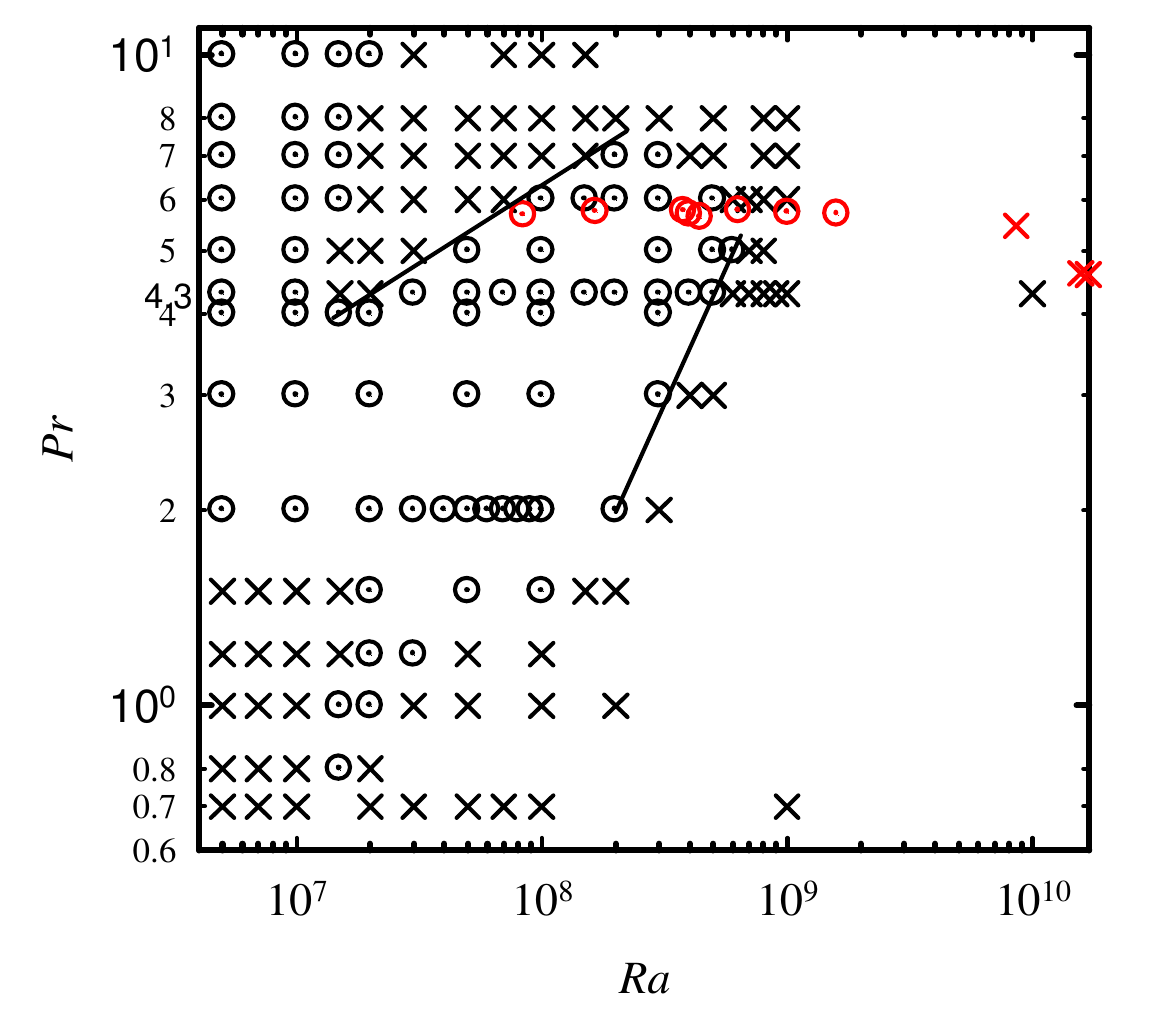} 
\caption
{(Color online) Phase diagrams in the Ra-Pr plane. Red symbols are from experiment and the black ones from DNS. Circles correspond to 
detected reversals [23], crosses to no detected reversals, in spite of excessive simulation (or
observation) time.
 The straight lines are guides to the eye; they have (from left to right) slopes 0.25 and 1.00.
} 
\label{fig:pd}
\end{figure}

These findings 
 led us to map out a considerable fraction of the Ra-Pr parameter space. The result is shown in figure \ref{fig:pd}, where the black symbols are from simulation and red symbols are from experiment. One sees a rather complicated structure. But given the limited amount of data, experiment and simulation are in general agreement, especially considering the fact that the simulations are
  for the true 2D case whereas the experiments run in a quasi-2D cell. It should again
 be pointed out that the experimental data point with the 
 highest Ra ($=1.6\times 10^9$)  that
 still shows a reversal has an extremely low reversal rate (0.5/day), 
which corresponds to waiting for $\sim 5500$
 large scale turn over time for a single reversal to occur. 

From figure \ref{fig:pd} we conclude that 
not only for too large Ra (as compared to above case of figure 1 with $Ra = 10^8$ and $Pr=4.3$) the reversals do not occur any more,
but also for 
too large or too small  Pr.  
How to physically understand 
 this complicated behavior? The key towards an understanding lies, from our point of view, in the role of the corner flows, and is based on a detailed observation of many movies at various Ra and Pr (see accompanying material).
 As stated above, the corner flow rolls are fed by plumes detaching from the plates' boundary layers. 
For too small Pr (i.e., too large thermal diffusivity) the thermal energy they carry is lost through thermal diffusion.
On the other hand, for too large Pr (i.e., too large kinematic 
viscosity) the thermal BL is nested in the kinematic BL and the thermal coupling of the corner flow towards the thermal BL is 
hindered. In both cases
the buildup of the corner flow and thus the reversals are suppressed. 
The situation is similar to the one in rotating RB, where Ekman vortices form, sucking heat out of the thermal BL and enhancing
the heat flux: Also here there is an optimal $Pr\approx 10$ for which the Nusselt number is maximal, and for larger or smaller 
Pr the very same above mechanisms hinder efficient heat transport \cite{ste10a}.

We now quantify this argument. 
The heat influx feeding the corner flow scales as
$J_{in} \sim \kappa \Delta H^{-1} Nu$. The outflux of thermal and kinetic energy
 is either of diffusive or of convective origin. We model it as $J_{out} = J_{out}^{diff} + J_{out}^{conv}$. Flow reversal is prevented if $J_{out} > J_{in}$. The convective outflux, which is dominant for large Pr, is modeled by $J_{out}^{conv} \sim J_{in} \cdot \lambda_u/\lambda_\theta \sim \kappa \Delta H^{-1} Nu^2 /\sqrt{Re}$. The diffusive outflux is $J_{out}^{diff} \sim \kappa_t \Delta H^{-1} Nu$ with some effective, turbulent thermal diffusivity $\kappa_t = \nu_t /Pr \sim Pr^{-1} U^4/\epsilon \sim \kappa Pr^2 Re^4 / (Nu Ra)$, where we have assumed $Pr_t \approx Pr$ and employed the $k$-$\epsilon$-model \cite{pop00} for the  turbulent viscosity $\nu_t$.

For dominant diffusive outflux (thus low Pr), suppression of reversals occurs at $\kappa_t \ge \kappa$. The threshold is given by the scaling relation $Nu Ra \sim Pr^2 Re^4$. Inserting $Nu(Ra,Pr) $ and $Re(Ra,Pr)$ either from experiment or from the unifying theory of refs.\ 
%\cite{gro00,gro01,gro02}, 
\cite{gro00}, 
one obtains a relation between the critical Prandtl number $Pr_{crit}$ and the critical Rayleigh number $Ra_{crit}$ at which reversals stop. Depending on whether regime $I_u$, $II_u$, or $IV_u$ of refs.\ 
%\cite{gro00,gro01,gro02} 
\cite{gro00}
is dominant, we obtain $Pr_{crit} \sim Ra_{crit}^{\gamma }$, with $\gamma = 3/5$ or $2/3$, respectively, which correctly reflects the trend in Fig.~\ref{fig:pd}. 

For large $Pr$ the convective outflux will be dominant. Here the threshold condition is $\kappa \sim \kappa Nu/\sqrt{Re}$, which with $Nu(Ra,Pr)$ and $Re(Ra,Pr)$ in regime $I_u$ of refs.\ 
%\cite{gro00,gro01,gro02} 
\cite{gro00}
leads to an Ra-independent $Pr_{crit}$, beyond which no reversals are possible. The reality of figure \ref{fig:pd} is clearly more complicated
\cite{footnote}
%{\footnote{For small Ra and large Pr (upper left corner of the phase diagram Fig.\ \ref{fig:pd}) the flow is plume dominated, has a very large coherence length, and no developed rolls exist. However, the angular momentum has zeros.}}, 
but at least the general trends are consistent with this explanation.

Finally, we note that 
we also performed experiments and simulations for  $\Gamma = 0.85$. 
Even for this relatively small change in $\Gamma$ the overall flow dynamics is very different and 
much more complex as compared to 
 the case of $\Gamma = 1$. 
Just as the important role the corner flows play for reversals, 
this finding demonstrates the strong effect  of the cell geometry  
on  the 
overall flow dynamics in the $\Gamma = \cal{O}(1) $ regime. In full 3D geometries, it may be less pronounced, but it certainly is present, too, see also ref.\ 
\cite{day01}. 
It remains remarkably that the rich structure in the (Ra, Pr, $\Gamma$) parameter space for reversals 
is hardly  reflected in Nu and Re.

\noindent
{\it Acknowledgements:} We thank E.\ Calzavarini for discussions and co-developing the code. Moreover, 
we acknowledge support 
by the Research
Grants Council of Hong Kong SAR (Nos. CUHK403806 and 403807) (R.N., S.Q.Z., H.D.X., K.Q.X.), 
and by the research programme of FOM, which
is financially supported by NWO (R.J.A.M.S., D.L.).

%\bibliographystyle{/Users/lohsed/Documents/papers/sty-files/prsty}
%\bibliography{/Users/lohsed/Documents/papers/bib-files/literatur}

%\bibliographystyle{prsty}
%\bibliography{literatur}

\begin{thebibliography}{10}
\bibitem{footnote}
For small Ra and large Pr (upper left corner of the phase diagram Fig.\ \ref{fig:pd}) the flow is plume dominated, has a very large coherence length, and no developed rolls exist. However, the angular momentum has zeros.
\end{thebibliography}

\end{document}